# A conductive topological insulator with colossal spin Hall effect for ultra-low power spin-orbit-torque switching

Nguyen Huynh Duy Khang[1], Yugo Ueda[1], and Pham Nam Hai[1,2*]

[1]*Department of Electrical and Electronic Engineering, Tokyo Institute of Technology,*

*2-12-1 Ookayama, Meguro, Tokyo 152-0033, Japan*

[2] *Center for Spintronics Research Network (CSRN), The University of Tokyo,*

*7-3-1 Hongo, Bunkyo, Tokyo 113-8656, Japan*

*Corresponding author: pham.n.ab@m.titech.ac.jp

**Spin-orbit-torque (SOT) switching using the spin Hall effect (SHE)[1,2] in heavy metals[3,4,5,6] and topological insulators (TIs) [7,8] has great potential for ultra-low power magnetoresistive random-access memory (MRAM). To be competitive with conventional spin-transfer-torque (STT) switching, [9] a pure spin current source with large spin Hall angle ($\theta_{SH} > 1$) and high electrical conductivity ($\sigma > 10^5$ $\Omega^{-1}$m$^{-1}$) is required. Here, we demonstrate such a pure spin current source: BiSb thin films with $\sigma \sim 2.5\times10^5$ $\Omega^{-1}$m$^{-1}$, $\theta_{SH} \sim 52$, and spin Hall conductivity $\sigma_{SH} \sim 1.3\times10^7 \frac{\hbar}{2e}\Omega^{-1}$m$^{-1}$ at room temperature. We show that BiSb thin films can generate a colossal spin-orbit field of 2770 Oe/(MA/cm$^2$) and a critical switching current density as low as 1.5 MA/cm$^2$ in Bi$_{0.9}$Sb$_{0.1}$ / MnGa bi-layers. BiSb is the best candidate for the first industrial application of topological insulators.**

MRAM is an emerging non-volatile memory technology that is gaining steam in various applications with no idling power consumption, un-limited endurance, and fast read/write time. Commercial MRAM uses STT switching method to write data to magnetic tunnel junctions (MTJs). In this method, a spin-polarized charge current $I_C$ is injected from the reference layer to the recording layer of a MTJ, generating a spin current $I_S = (\hbar/2e)PI_C$, where $\hbar$ is the reduced Plank constant, $e$ is the elementary charge, and $P$ is the spin-polarisation of the reference layer. This limits the amount of spin current generated by a given $I_C$ and makes it difficult to reduce the writing current. Indeed, the writing current and writing energy of MRAM is worse than conventional volatile memory technologies by one order of magnitude. In addition, high writing current requires large driving transistors, which makes it difficult to increase the bit density of MRAM. Recently, SOT switching using the giant SHE in heavy metals and TIs has attracted much attention as an alternating writing method for MRAM. In SOT switching, a spin Hall layer is in contact with the recording magnetic layer. A charge current flowing in the spin Hall layer can generate a pure spin current that exerts a spin torque on the recording layer. The spin current generated by this way is given by $I_S = (\hbar/2e)(L/t)\theta_{SH} I_C$, where $L$ is the MTJ size and $t$ the thickness of the spin Hall layer. Since $L/t$ is about 10 in realistic MTJs, SOT switching may be more effective than STT switching if $\theta_{SH} > 1$. Such giant $\theta_{SH}$ has recently been observed in several TIs, such as $Bi_2Se_3$ ($\theta_{SH}$ = 2-3.5 at 300 K) (Ref. 7), $Bi_xSe_{1-x}$ ($\theta_{SH}$ = 18.8 at 300 K) (Ref. 10), and $(Bi_{0.5}Sb_{0.5})_2Te_3$ ($\theta_{SH}$ = 150-450 at cryogenic temperature) (Ref. 8). Because of their large band gap and low carrier mobility, however, the conductivity $\sigma$ of those TIs is at most $6\times10^4$ $\Omega^{-1}m^{-1}$, which is much lower than that of typical ferromagnetic and non-magnetic metals used in realistic MRAM (for example, $\sigma \sim 6\times10^5$ $\Omega^{-1}m^{-1}$ for CoFeB and $5\times10^5$ $\Omega^{-1}m^{-1}$ for Ta). This causes a serious problem for TIs as spin Hall materials: when a TI layer is attached to a metallic ferromagnet and a Ta capping/buffering layer, most of the charge current is shunted through those metallic layers and does not contribute to generation of pure spin current in the TI layer. To be used in realistic SOT-MRAM, the spin Hall material must have both large $\theta_{SH}$ and high $\sigma$. Here, we face a dilemma: heavy metals (such as Pt, Ta, and W) have high $\sigma$ but small $\theta_{SH}$ (~ 0.1), while TIs have large $\theta_{SH}$ but low $\sigma$. There is no spin Hall material so far that can satisfy both conditions simultaneously.

Here, we demonstrate that BiSb can satisfy both conditions. $Bi_{1-x}Sb_x$ ($0.07 \leq x \leq 0.22$) is a very narrow-gap TI with strong spin orbit coupling. [11] Its non-trivial topologically protected surface states have been confirmed by the angle-resolved photoemission spectroscopy (ARPES)[12,13,14] and magneto-transport measurements[15,16]. Despite being the first 3 dimensional TI identified by ARPES, BiSb has received little attention due to its very small bulk band gap (< 20 meV) and complex Fermi surfaces. However, BiSb is very attractive as a spin Hall material: thanks to its high carrier mobility (~$10^4$ cmV$^{-1}$s$^{-1}$), its bulk conductivity is as high as $4 \times 10^5 \sim 6.4 \times 10^5$ $\Omega^{-1}$m$^{-1}$, compatible to other metallic materials used in realistic MRAM. Recently, we have developed epitaxial growth technique for high quality BiSb thin films using molecular beam epitaxy (MBE),[17] and obtained $\sigma = 4 \times 10^5 \sim 6 \times 10^5$ $\Omega^{-1}$m$^{-1}$ for BiSb thin films thicker than 80 nm, and $\sigma = 1 \times 10^5 \sim 4 \times 10^5$ $\Omega^{-1}$m$^{-1}$ (average $\sigma \sim 2.5 \times 10^5$ $\Omega^{-1}$m$^{-1}$) for BiSb thin films thinner than 25 nm in the TI region ($0.07 \leq x \leq 0.22$) (see Supplementary Information). In this work, we investigate the performance of thin BiSb films as a pure spin current source in various BiSb / MnGa bi-layers. We observe a colossal spin Hall effect of $\theta_{SH} \sim 52$ and $\sigma_{SH} \sim 1.3 \times 10^7 \frac{\hbar}{2e} \Omega^{-1}$m$^{-1}$ at room temperature, making BiSb the best candidate for the spin current source in ultra-low power SOT-MRAM.

The BiSb / MnGa bi-layers in this work were deposited on semi-insulating GaAs(001) substrates with crystallographic orientation of $Bi_{0.9}Sb_{0.1}$(012)//MnGa(001)//GaAs(001) by MBE. By optimizing the growth condition, we can obtain high quality BiSb / MnGa bi-layers with atomically smooth interface, despite the large lattice mismatch between MnGa(001) and BiSb(012) (see Method and Supplementary Information for details). Here, we choose MnGa as a model ferromagnet because its large perpendicular magnetic anisotropy (> 10 Merg/cc), large uniaxial anisotropy field (> 40 kOe), large coercive force (> 1.5 kOe), and high conductivity ($5\times10^5$ $\Omega^{-1}$m$^{-1}$) represent well those would be used in futuristic ultra-high density SOT-MRAM. Furthermore, by adjusting the Mn composition, we are able to fabricate $Mn_{0.6}Ga_{0.4}$ layers with perfect perpendicular magnetisation, and $Mn_{0.45}Ga_{0.55}$ layers with titling magnetisation which are convenient for evaluation of the spin-orbit field generated by the BiSb layer. Figure 1(a) shows the schematic $Bi_{0.9}Sb_{0.1}$ (10) / $Mn_{0.6}Ga_{0.4}$ (3) (thickness in nm) bi-layer and the coordinate system used in this work. An electric current was applied along the $x$ direction, while an external magnetic field $H_{ext}$ was applied in the $z$-$x$ plane at an angle $\theta$ with respect to the $z$ axis. Figure 1(c) shows the out-of-plane magnetisation curve of the bi-layer measured at various temperatures. The magnetisation of the 3 nm-thick $Mn_{0.6}Ga_{0.4}$ layer in this bi-layer is the same as that of high quality stand-alone $Mn_{0.6}Ga_{0.4}$ thin films grown on GaAs(001), indicating that there is no interfacial magnetic dead layer. The bi-layers are then patterned into 100 μm-long and 50 μm-wide Hall bars for transport measurements, as one shown in Fig. 1(b). The Hall bars are mounted inside a liquid nitrogen cryostat which acts as a heat sink to minimize the effect of Joule heating. Figure 1(d) shows the Hall resistance $R_H$ of a Hall bar under a perpendicular magnetic field ($\theta = 0$) at 150 K, measured with various current density $J = 0.2 - 15.4\times10^5$ A/cm$^2$. Here, $J$ is the nominal current density of the Hall bar, defined as $I/[w*(t_{BiSb} + t_{MnGa})]$, where $I$ is the applied current (10 mA maximum), $w = 50$ μm is the width of the Hall bar, $t_{BiSb} = 10$ nm and $t_{MnGa} = 3$ nm are the thickness of the BiSb and MnGa layer, respectively. The Hall resistance is dominated by the anomalous Hall effect which reveals the perpendicular magnetisation component of the $Mn_{0.6}Ga_{0.4}$ layer. We observed a systematic reduction of the coercive force $H_c$ of the $Mn_{0.6}Ga_{0.4}$ layer when $J$ increased. The reduction $\Delta H_c$ is as large as 2 kOe at a modest $J = 15.4 \times 10^5$ A/cm$^2$. To make sure that this reduction of $H_c$ is not the result of Joule heating of the $Mn_{0.6}Ga_{0.4}$ layer, we prepared a Hall bar from a stand-alone $Mn_{0.6}Ga_{0.4}$ (3) layer, and measured its Hall resistance under various $J$. As seen in Fig. 1(e), we observed no clear change of $H_c$ up to $J = 46.7 \times 10^5$ A/cm$^2$, which is 3 time higher than the highest current density applied in Fig. 1(d). This indicates that the reduction of $H_c$ shown in Fig. 1(d) is not the result of Joule heating. Indeed, for the Joule heating to be the origin of the large $\Delta H_c = 2$ kOe, the $Mn_{0.6}Ga_{0.4}$ layer temperature had to increase from 150 K to 250 K (see Fig. 1(c)), which is unlikely given the very small $J$ used in our experiments, comparing with $J \sim 10^7$ - $10^8$ A/cm$^2$ used in previous studies using heavy metals as the spin Hall layer. Furthermore, we observed a significant reduction of the remanent $R_H$ at $J = 15.4 \times 10^5$ A/cm$^2$, which was not observed in the magnetisation curve at 250 K. Therefore, we conclude that the large reduction of $H_c$ is the result of an in-plane spin-orbit field generated by spin current injection from the BiSb layer. Figure 1(f) shows $\Delta H_c$ as a function of the current density $J_{BiSb}$ in the BiSb layer. Here, $J_{BiSb}$ is calculated using $\sigma_{BiSb} = 2.5 \times 10^5$ $\Omega^{-1}$m$^{-1}$ and $\sigma_{MnGa} = 5 \times 10^5$ $\Omega^{-1}$m$^{-1}$. The gradient $\Delta H_c/\Delta J_{BiSb}$ is as large as 3700 Oe/(MA/cm$^2$) at high $J_{BiSb}$ (dashed line in Fig. 1(f)).

In order to quantitatively evaluate the spin Hall angle of BiSb, we prepared another bi-layer $Bi_{0.9}Sb_{0.1}$ (10) / $Mn_{0.45}Ga_{0.55}$ (3) with tilting magnetic domains, as shown in the inset of Fig. 2(a). The blue and the red curves in Fig. 2(a) show the in-plane and out-of-plane magnetisation curve of the as-grown bi-layer at room temperature. Because of the tilting magnetisation, the out-of-plain magnetisation does not saturate until a perpendicular $H_{ext} \sim 20$ kOe is applied. The in-plane magnetisation shows small remanence, but rapidly increases up to an in-plane $H_{ext} \sim 4$ kG, then slowly increases and saturates at an in-plane $H_{ext} \sim 50$ kOe. The initial rise of the in-plane magnetisation at the small in-plane $H_{ext}$ indicates reorientation of the magnetic domains toward the in-plane $H_{ext}$ direction, while the slow increase at the in-plane $H_{ext} > 4$ kG reflects their tilting toward the horizontal direction. This unique magnetic behaviors of the $Mn_{0.45}Ga_{0.55}$ layer allow us to directly evaluate $H_{so}$ as discussed below. Figure 2(b) shows the $R_H$ of a 100 μm × 50 μm Hall bar made from the bi-layer in Fig. 2(a) under nearly perpendicular $H_{ext}$ at room temperature. $H_{ext}$ is slightly tilted at $\theta \sim 2°$ so that its in-plane component at 8 kG can reorientate those tilting magnetic domains toward the $x$ direction and induces an in-plane $M_x$ component of the $Mn_{0.45}Ga_{0.55}$ layer (see Fig. 1(a) and the inset in Fig. 2(a)). Since the direction of the damping-like spin-orbit field $H_{so}$ is given by $-\boldsymbol{\sigma} \times \boldsymbol{m}$, where $\boldsymbol{\sigma}$ is the spin polarisation unit vector of the spin current along the $y$ direction and $\boldsymbol{m}$ is the unit vector of the magnetisation direction, there is an out-of-plane component $H_{so-z}$ of the spin-orbit field associating with $M_x$. The existence of such $H_{so-z}$ component can be seen in Fig. 2(b); the remanence and coercive force of the $R_H$ hysteresis systematically increase with increasing positive $J$. This clearly cannot be explained by Joule heating which can only reduce but not enhance the remanent $R_H$ and coercive force. For a reference, we show the $R_H$ hysteresis with reduced remanence and coercive force at a negative $J = -7.7 \times 10^5$ A/cm$^2$. We also found that the field-like spin-orbit field is negligible (See Supplementary Information). To further confirm the effect of $H_{so-z}$, we measured $R_H$ under an in-plane $H_{ext}$ ($\theta=90°$). Figure 2(c) and 2(d) show the $R_H$-$H_{ext}$ ($\theta=90°$) hysteresis at $J = 13.8 \times 10^5$ A/cm$^2$ and $J = -13.8 \times 10^5$ ×10$^5$ A/cm$^2$, respectively. We observed that $R_H$ rapidly increases up to $H_{ext} \sim 4$ kG, and then gradually decreases at $H_{ext} > 4$ kG. This behavior is consistent with the in-plane magnetisation curve shown in Fig. 2(a); at $H_{ext} = 0~4$ kG, the magnetic domains are rapidly reoriented toward the $x$-directions so that $H_{so-z}$ and $R_H$ reach their maximum value at 4 kG, after that the magnetisation begins tilting toward the $x$ direction and $R_H$ decreases. Furthermore, we observed that $H_{so-z}$ and $R_H$ switch its directions when direction of $M_x$ or $J$ is switched, consistent with the behavior of SOT. These data also show that $\theta_{SH}$ of BiSb has the same sign as those of $(Bi_{0.5}Sb_{0.5})_2Te_3$ (Ref. 8), $Bi_2Se_3$ (Ref. 7) and Pt (Ref. 5). Next, we evaluate the spin Hall angle from data in Fig. 2(b). We note that at $H_{ext} = -H_c(J)$ (the coercive force at a given $J$), the net perpendicular magnetisation component is zero, so there is only the in-plane magnetisation component. The situation at $H_{ext} = -H_c(J)$ is shown in the inset of Fig. 2(b). At this point, $\boldsymbol{H}_{so}$ points to the $z$ direction and counters $\boldsymbol{H}_{ext}$, meaning that the shift of the coercive $\Delta H_c$ is nothing other than $H_{so}$. Therefore, we obtain $H_{so} = 3.1$ kOe at $J = 13.8 \times 10^5$ A/cm$^2$ ($J_{BiSb} = 11.2 \times 10^5$ A/cm$^2$), yielding a colossal spin orbit field of $H_{so}/J_{BiSb} = 2770$ Oe/(MA/cm$^2$) at room temperature. This value is much larger than those of heavy metals and TIs reported before, such as Ta ~ 6.8 Oe/(MA/cm$^2$) in Fe (1.1) / Ta (7.2), Pt ~ 2.9 Oe/(MA/cm$^2$) in Fe (0.5) / Pt (2.3) (Ref. 18), and $Bi_xSe_{1-x}$ ~100 Oe/(MA/cm$^2$) in CoFeB (5) / $Bi_xSe_{1-x}$ (4) (Ref. 10). The room-temperature spin Hall angle, calculated by

$\theta_{SH} = \frac{2e}{\hbar} M_{MnGa} t_{MnGa} \frac{H_{SO}}{J_{BiSb}}$, is 52, which is the highest value reported so far.

In order to demonstrate SOT switching with ultra-low current density using the colossal spin Hal effect of BiSb, we prepare a 100 μm × 50 μm Hall bar of a $Bi_{0.9}Sb_{0.1}$ (5) / $Mn_{0.45}Ga_{0.55}$ (3) bi-layer. Figure 3(a) and 3(b) demonstrate the SOT switching of the MnGa layer when applying 100 ms pulse currents to the Hall bar and an in-plane $H_{ext} = + 3.5$ kOe and -3.5 kOe, respectively. We observed clear switching at an average critical current density of $J = 1.5 \times 10^6$ A/cm$^2$ ($J_{BiSb} = 1.1 \times 10^6$ A/cm$^2$). Here, the critical current density is defined at which the Hall resistance changes sign. Furthermore, the switching direction is reversed when the in-plane $H_{ext}$ direction is reversed, consistent with the behavior of SOT switching. The observed critical current density is much smaller than those of Ta (5) / MnGa (3) ($J = 1.1 \times 10^8$ A/cm$^2$),[19] IrMn (4) / MnGa (3) ($J = 1.5 \times 10^8$ A/cm$^2$),[20] and Pt (2) / MnGa (2.5) ($J = 5.0 \times 10^7$ A/cm$^2$).[21]

BiSb has many characteristics that make it the best candidate for the pure spin current source in SOT-MRAM. Its conductivity is comparable to typical ferromagnetic and non-magnetic metals used in MRAM, while its spin Hall angle is as large as 52 at room temperature. It can be grown well on ferromagnetic metals without creating a magnetic dead layer. Table 1 summarises $\sigma$, $\theta_{SH}$, and $\sigma_{SH}$ of several heavy metals and TIs at room temperature. In term of $\sigma_{SH}$, which is considered as the figure of merit for spin Hall materials, BiSb outperforms the nearest competitor (Pt) by a factor of 30. Using BiSb, we have demonstrated both colossal spin-orbit field of 2770 Oe/(MA/cm$^2$) and SOT switching at a low critical current density of $1.5 \times 10^6$ A/cm$^2$, even though the MnGa ferromagnet used in our bi-layer has higher PMA by one order of magnitude than those used in previous room-temperature SOT switching experiments in Bi$_2$Se$_3$/CoTb (Ref. 24) and Bi$_x$Se$_{1-x}$/CoFeB (Ref.10). These double check the colossal spin Hall effect of BiSb. Here, we estimate the critical switching current for a BiSb-based SOT-MRAM with size of 37 nm and a 5 nm-thick BiSb layer as the spin current source. Since the magnetic properties of the MnGa layer used in our work ($M_St$ ~ 750 (emu/cc)nm, coercive force ~ 1.6 kOe, and anisotropic magnetic field ~ 50 kOe) are close to those used in realistic perpendicular MTJs, we assume the same critical current density of $1.5 \times 10^6$ A/cm$^2$. This yields a critical switching current of 2.2 µA, which is one order of magnitude smaller than that of STT-MRAM at the same size (24 µA).[25] Therefore, the switching power can be reduced by at least one order of magnitude. Furthermore, since SOT-MRAM can be switched one order of magnitude faster than STT-MRAM,[26] the switching energy can be reduced by at least two orders of magnitude. That means BiSb-based SOT-MRAM can be very competitive to even SRAM, and is the suitable ultra-low-power memory for internet-of-thing applications. As over ten years have passed from the proposals[27,28] and the first realisation of TIs,[29] BiSb emerges as the best candidate for the first industrial application of TIs.

The observation of colossal spin Hall effect in BiSb leaves many open questions about the physics of spin Hall effect in TIs. Even though stronger spin-orbit interaction and larger spin Hall effect are expected for BiSb than $Bi_3Se_2$ or $(Bi,Sb)_3Te_2$ thanks to its very small band gap, the observation of colossal $\sigma_{SH} = 1.3 \times 10^7 \frac{\hbar}{2e} \Omega^{-1} m^{-1}$ is unexpected. Because of the very small band gap, the electric current flows in both the surface and the bulk of BiSb, thus contribution of both surface and bulk spin Hall effect should be taken into account. However, first-principle calculation of the bulk intrinsic spin Hall effect of BiSb yields a maximum value of $4.9 \times 10^4 \frac{\hbar}{2e} \Omega^{-1} m^{-1}$ when the Fermi level is in the band gap, which can account for only 0.37% of the observed value (although very large bulk intrinsic $\theta_{SH} = \frac{2e}{\hbar} \frac{\sigma_{SH}}{\sigma}$ is possible from the calculation since the bulk $\sigma \sim 0$ at zero temperature).[30] Furthermore, the bulk extrinsic mechanism such as side-jump or skew-scattering is unlikely the main mechanism given the high carrier mobility and low scattering rate in BiSb alloys. We therefore suggest that the topological surface states of BiSb(012) are mainly responsible for the observed colossal spin Hall effect in BiSb. Detailed observation of the BiSb(012) surface states by ARPES is needed for unlocking the secret of the colossal spin Hall effect in BiSb.

**Acknowledgements**

This work is supported by Grant-in-Aid for Challenging Exploratory Research (No. 16K14228), and Nanotechnology platform 12025014 (F-17-IT-0011) from MEXT. The authors gratefully thank to H. Iida, and R. C. Roca for their help in XRD and SQUID measurements. We also thank the Material Analysis Division and Laboratory for Future Interdisciplinary Research of Science and Technology at Tokyo Institute of Technology, and M. Tanaka Laboratory at University of Tokyo for their technical supports.

**Methods**

**MBE growth.** The thin films were grown on semi-insulating GaAs(001) substrates by using ultrahigh vacuum MBE system. After removing the surface oxide layer of the GaAs substrate at 580°C, a 100 nm-thick GaAs buffer layer was grown to obtain an atomically smooth surface. The substrate was cooled down to room temperature for deposition of a 4 monolayer Mn-Ga-Mn-Ga template. Then, the substrate was heated up to 250°C for deposition of a MnGa thin film with the total thickness of 3 nm. A $Bi_{0.1}Sb_{0.9}$ layer was grown on top of the MnGa layer at 200°C with the rate of 1 nm/min. Finally, the samples were cooled down again to room temperature for deposition of an As thin cap layer. The growth process was monitored *in situ* by reflection high energy electron diffraction.

**Hall bar fabrication.** The samples were patterned into 100 µm × 50 µm Halls bar by optical lithography and Ar ion milling. An 150 nm-thick Au with a 5 nm-thick Cr adhesion layer were deposited as electrodes by electron beam evaporation.

**SOT measurements.** A Keithley 2400 Sourcemeter was used as the current source for DC and pulse measurements. For the pulse measurements, a 100 ms pulse was first applied, then a small

DC current of 0.1 mA was applied to measure the Hall resistance. The Hall voltage was measured using an ADCMT 7461A Digital Mutilmeter or a Keithley 2182A Nanovolmeter. The Hall bars were mounted inside a cryostat equipped with a rotatable electromagnet. The cryostat was not cooled for room temperature measurements, but was cooled by liquid nitrogen for low temperature measurements.

**References**

[1] Kato, Y., Myers, R. C., Gossard, A. C. & Awschalom, D. D. Observation of the Spin Hall Effect in Semiconductors. *Science* **306**, 1910–1913 (2004).

[2] Wunderlich, Kaestner, J. B., Sinova, J. & Jungwirth, T. Experimental Observation of the Spin-Hall Effect in a Two-Dimensional Spin-Orbit Coupled Semiconductor System. *Phys. Rev. Lett.* **94**, 047204 (2005).

[3] Miron, I. M. *et al*. Perpendicular switching of a single ferromagnetic layer induced by in-plane current injection, *Nature* **476**, 189-193 (2011).

[4] Liu, L., Pai, C. F., Li, Y., Tseng, H. W., Ralph, D. C. & Buhrman, R. A. Spin-torque switching with the giant spin Hall effect of Tantalum. *Science.* **336**, 555-558 (2012).

[5] Liu, L., Lee, O. J., Gudmundsen, T. J., Ralph, D. C. & Buhrmam, R. A. Current-induced switching of perpendicularly magnetized magnetic layers using spin torque from the spin Hall effect. *Phys. Rev. Lett.* **109**, 096602 (2012).

[6] Hao, Q. & Xiao, G. Giant spin Hall effect and switching induced by spin-transfer torque in a W/Co$_{40}$Fe$_{40}$B$_{20}$/MgO structure with perpendicular magnetic anisotropy. *Phys. Rev. Appl.* **3**, 034009 (2015).

[7] Melnik, A. R. *et al*. Spin-transfer torque generated by a topological insulator. *Nature* **511**, 449-451 (2014).

[8] Fan, Y. *et al*. Magnetisation switching through giant spin-orbit torque in a magnetically doped topological insulator heterostructure. *Nature Mater.* **13**, 699-704 (2014).

[9] Slonczewski, J. C. Current-driven excitation of magnetic multilayers. *J. Magn. Magn. Matter.* **159**, L1-L7 (1996).

[10] Mahendra, DC. *et al.* Room-temperature perpendicular magnetization switching through giant spin-orbit torque from sputtered BixSe(1-x) topological insulator material. Preprint at *arXiv*: 1703.03822v1.

[11] Teo, J. C. Y., Fu, L. & Kane, C. L. Surface states and topological invariants in three-dimensional topological insulators: Application to Bi$_{1-x}$Sb$_x$. *Phys. Rev. B* **78**, 045426 (2008).

[12] Hsieh, D. *et al*. A topological Dirac insulator in a quantum spin Hall phase. *Nature* **452**, 970-974 (2008).

[13] Hirahara, T. *et al*. Topological metal at surface of an ultrathin $Bi_{1-x}Sb_x$ alloy film. *Phys. Rev. B*. **81**, 165422 (2010).

[14] Nishide, A. *et al.* Direct mapping of the spin-filtered surface bands of a three-dimensional quantum spin Hall insulator. *Phys. Rev. B* **81**, 041309(R) (2010).

[15] Taskin, A. A & Ando, Y. Quantum oscillations in a topological insulator $Bi_{1-x}Sb_x$. *Phys. Rev. B* **80**, 085303 (2009).

[16] Taskin, A. A. , Segawa, K. & Ando, Y. Oscillatory angular dependence of the magnetoresistance in a topological insulator $Bi_{1-x}Sb_x$. *Phys. Rev. B* **82**, 121302 (R) (2010).

[17] Ueda, Y., Khang, N. H. D., Yao, K. & Hai, P. N. Epitaxial growth and characterization of $Bi_{1-x}Sb_x$ spin Hall thin films on GaAs(111)A substrates. *Appl. Phys. Lett.* **110**, 062401 (2017).

[18] Kawaguchi, M. *et al*. Current-induced effective fields detected by magnetotransport measurements. *Appl. Phys. Exp.* **6**, 113002 (2013).

[19] Meng, K. K. *et al.* Modulated switching current density and spin-orbit torques in MnGa/Ta films with inserting ferromagnetic layers. *Sci. Rep.* **6**, 38375 (2016).

[20] Meng, K. K. *et al*. Anomalous Hall effect and spin-orbit torques in MnGa/IrMn films: Modification from strong spin Hall effect of the antiferromagnet. *Phys. Rev. B* **94**, 214413 (2016).

[21] Ranjbar, R., Suzuki, K. Z., Sasaki, Y., Bainsla, L. & Mizukami, S. Current-induced spin-orbit torque magnetisation switching in a MnGa/Pt film with a perpendicular magnetic anisotropy. *Jpn. J. Appl. Phys.* **55**, 120302 (2016).

[22] Pai, C.-F. *et al.* Spin transfer torque devices utilizing the giant spin Hall effect of tungsten. *Appl. Phys. Lett.* **101**, 122404 (2012).

[23] Liu, L., Moriyama, T., Ralph, D. C. & Buhrman R. A. Spin-Torque Ferromagnetic Resonance Induced by the Spin Hall Effect. *Phys. Rev. Lett.* **106**, 036601 (2011).

[24] Han, J. *et al*. Room temperature spin-orbit torque switching induced by a topological insulator. *Phys. Rev. Lett.* **119**, 077702 (2017).

[25] Thomas, L. Basic Principles and Challenges of STT-MRAM for Embedded Memory Applications. *MSST 2017 Mass Storage Conference* (2017).

[26] Lee, K.-S., Lee, S.-W., Min, B.-Ch. & Lee, K.-J. Threshold current for switching of a perpendicular magnetic layer induced by spin Hall effect. *Appl. Phys. Lett.* **102**, 112410 (2013).

[27] Kane, C. L. & Mele, E. J. Quantum spin Hall effect in graphene. *Phys. Rev. Lett.* **95**, 226801 (2005).

[28] Bernevig, B. A., Hughes, T. L. & Zhang, S.-C. Quantum spin Hall effect and topological phase transition in hgte quantum wells. *Science* **314**, 1757–1761 (2006).

[29] König, M. *et al*. Quantum spin Hall insulator state in HgTe quantum wells. *Science* **318**, 766–770 (2007).

[30] Sahin, C. & Flatté, M. E. Tunable giant spin Hall conductivities in a strong spin-orbit semimetal: $Bi_{1-x}Sb_x$. *Phys. Rev. Lett.* **114**, 107201 (2015).

**Table 1. Room-temperature spin Hall angle $\theta_{SH}$, conductivity $\sigma$, and spin Hall conductivity $\sigma_{SH}$ of several heavy metals and TIs.**

|  | $\theta_{SH}$ | $\sigma$ ($\Omega^{-1}m^{-1}$) | $\sigma_{SH}$ ($\frac{\hbar}{2e}\Omega^{-1}m^{-1}$) |
|---|---|---|---|
| β-Ta (Ref. 4) | 0.15 | $5.3\times10^5$ | $0.8\times10^5$ |
| β-W (Ref. 22) | 0.4 | $4.7\times10^5$ | $1.9\times10^5$ |
| Pt (Ref. 23) | 0.08 | $4.2\times10^6$ | $3.4\times10^5$ |
| $Bi_2Se_3$ (Ref. 7) | 2-3.5 | $5.7\times10^4$ | $1.1$-$2.0\times10^5$ |
| $Bi_xSe_{1-x}$ (Ref. 10) | 18.8 | $7.8\times10^3$ | $1.47\times10^5$ |
| **$Bi_{0.9}Sb_{0.1}$ (this work)** | **52** | **$2.5\times10^5$** | **$1.3\times10^7$** |

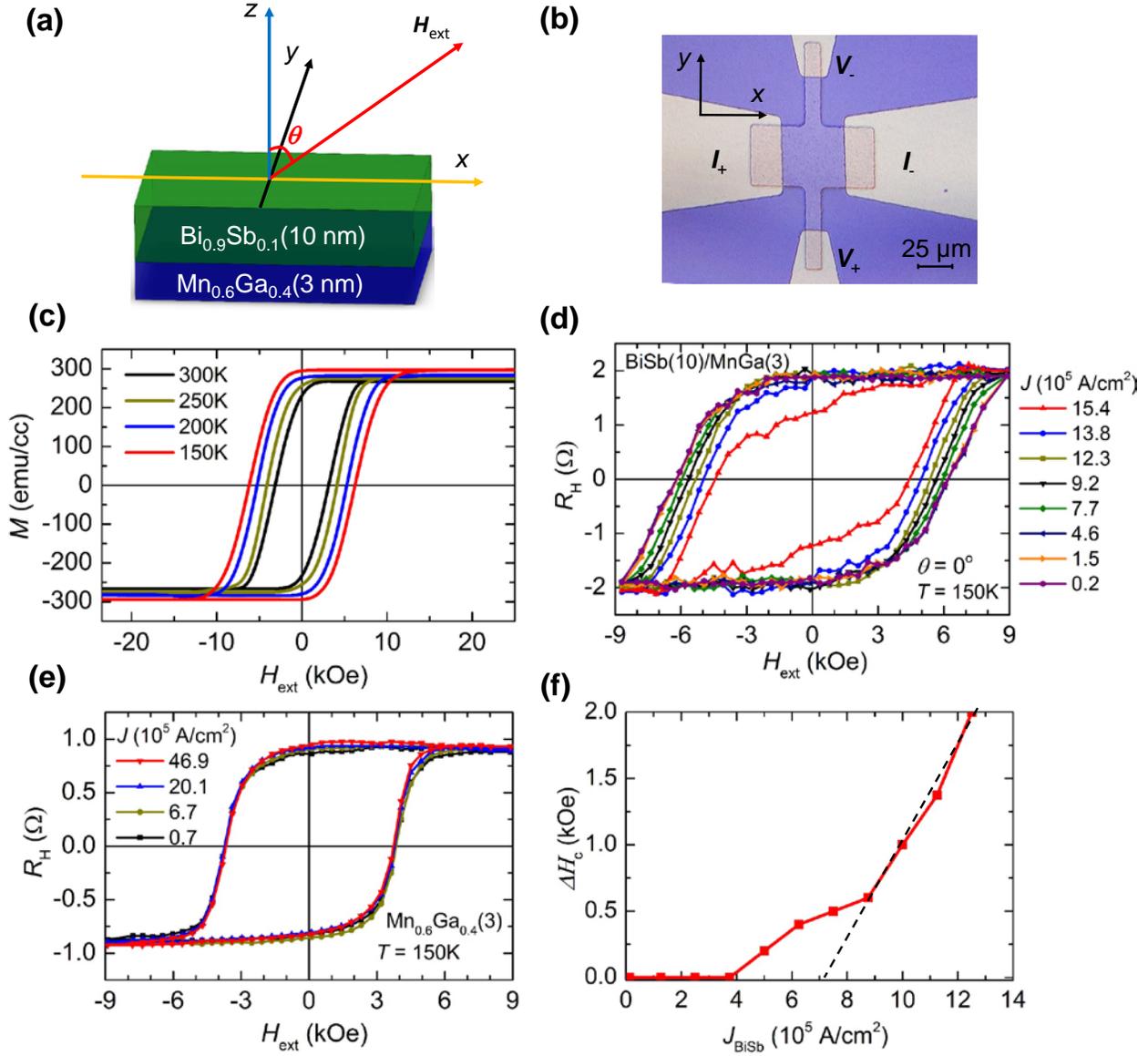

**Figure 1. Structure, magnetic properties, and SOT effect in the $Bi_{0.9}Sb_{0.1}$ (10) / $Mn_{0.6}Ga_{0.4}$ (3) bi-layer with perfect perpendicular magnetic anisotropy. a,** Schematic structure of the $Bi_{0.9}Sb_{0.1}$ (10) / $Mn_{0.6}Ga_{0.4}$ (3) (thickness in nm) bi-layer and the coordinate system used in this work. **b,** Micrograph of a 100 μm × 50 μm Hall bar device. **c,** Out-of-plane magnetisation curves of the bi-layer at different temperatures. **d,** Hall resistance $R_H$ of a Hall bar under a perpendicular magnetic field $H_{ext}$ ($\theta = 0$) at 150 K, measured with various current density $J = 0.2 - 15.4 \times 10^5$ A/cm$^2$. **e,** $R_H$ of a Hall bar of a stand-alone MnGa (3) layer. **f,** Reduction of the coercive force $\Delta H_c$ of the bi-layer as a function of the current density $J_{BiSb}$ in the BiSb layer.

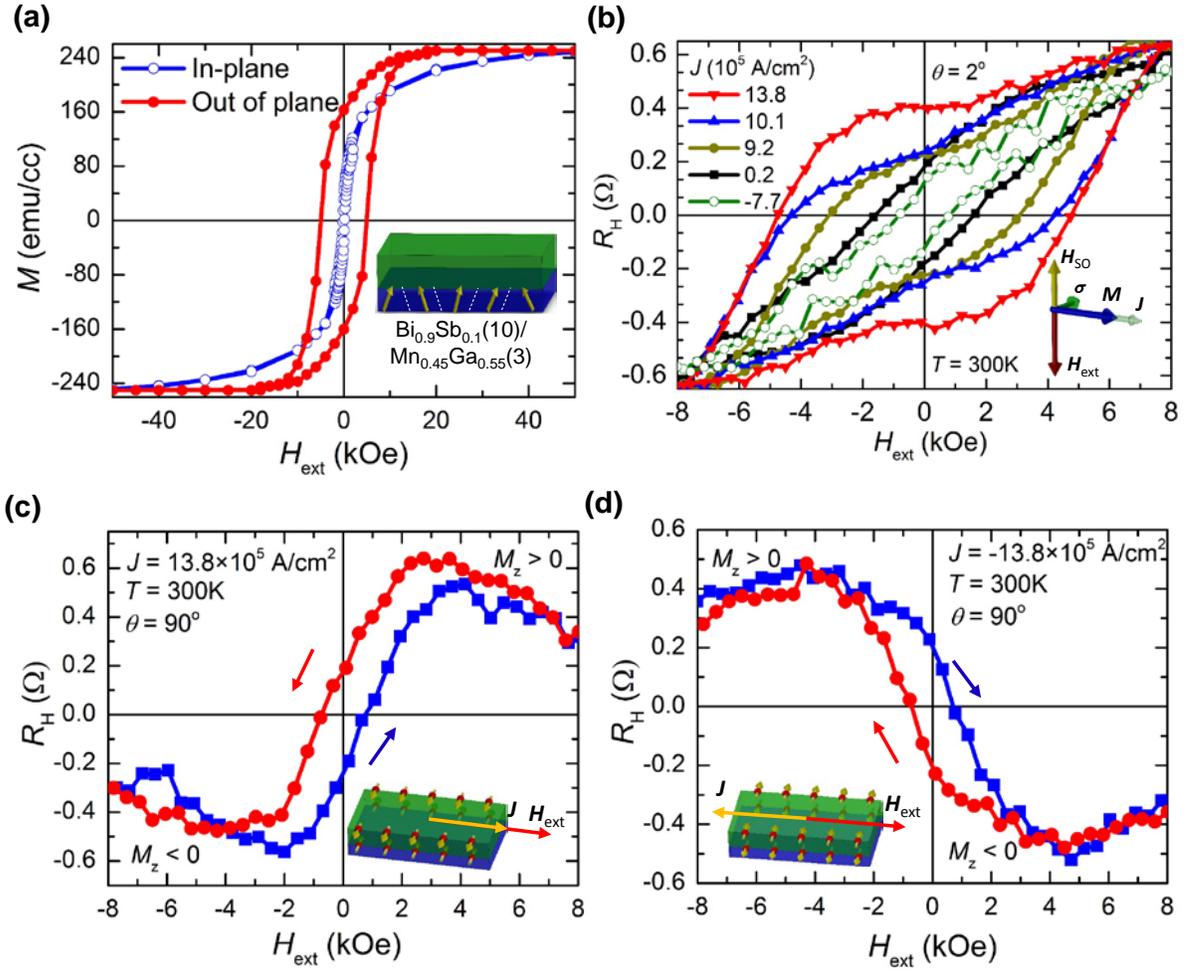

**Figure 2. Magnetic properties and SOT effect in the $Bi_{0.9}Sb_{0.1}$ (10) / $Mn_{0.45}Ga_{0.55}$ (3) bi-layer with tilting magnetisation. a,** In-plane (blue) and out-of-plane (red) magnetisation curve of the as-grown bi-layer. Inset shows the tilting magnetic domains of the $Mn_{0.45}Ga_{0.55}$ layer. **b,** $R_H$ of a 100 μm × 50 μm Hall bar of the bi-layer under a slightly tilting perpendicular magnetic field ($\theta = 2°$), measured with various current density $J = -7.7 – 13.8 \times 10^5$ A/cm². Inset shows the perpendicular spin-orbit field $\mathbf{H_{so}}$ acting on the magnetisation vector $\mathbf{M}$ at $H_{ext} = -H_c$ (coercive force) **c,d,** $R_H$ as a function of an in-plane $H_{ext}$ ($\theta = 90°$) at $J = +13.8 \times 10^5$ A/cm² and $-13.8 \times 10^5$ A/cm², respectively. All data in figures **a-d** are taken at 300 K.

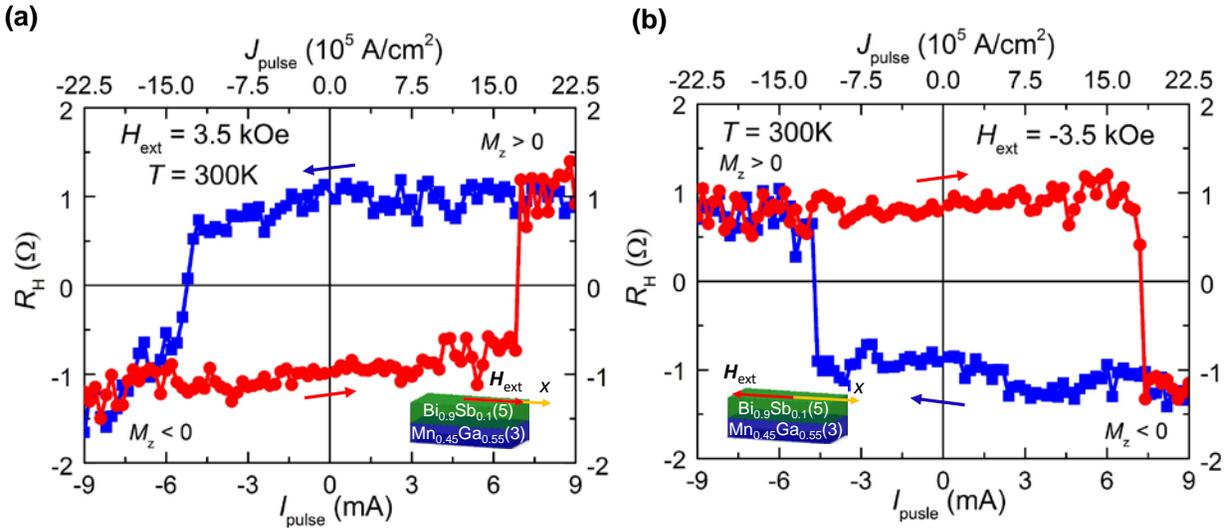

**Figure 3. Room-temperature current induced magnetization switching in the Bi$_{0.9}$Sb$_{0.1}$ (5) / Mn$_{0.45}$Ga$_{0.55}$ (3) bi-layer. a,b,** $R_H$ of a 100 μm × 50 μm Hall bar of the bi-layer under 100 ms pulse currents and an in-plane $H_{ext}$ of +3.5 kOe and -3.5 kOe, respectively.

# Supplementary Information

# A conductive topological insulator with colossal spin Hall effect for ultra-low power spin-orbit-torque switching

Nguyen Huynh Duy Khang[1], Yugo Ueda[1], and Pham Nam Hai[1,2*]

[1]*Department of Electrical and Electronic Engineering, Tokyo Institute of Technology,*

*2-12-1 Ookayama, Meguro, Tokyo 152-0033, Japan*

[2] *Center for Spintronics Research Network (CSRN), The University of Tokyo,*

*7-3-1 Hongo, Bunkyo, Tokyo 113-8656, Japan*

*Corresponding author: pham.n.ab@m.titech.ac.jp

## 1. Conductivity of BiSb thin films

Figure S1(a) shows the conductivity $\sigma$ of BiSb thin films at various thickness and Sb atomic concentration at 270 K. In the TI area ($7\% < x < 22\%$), $\sigma = 4\times10^5 \sim 6\times10^5$ $\Omega^{-1}$m$^{-1}$ for BiSb thin films thicker than 80 nm, which are the same as bulk values. For thinner films (< 60 nm), $\sigma$ is reduced to $1\times10^5 \sim 4\times10^5$ $\Omega^{-1}$m$^{-1}$ (average $\sigma \sim 2.5\times10^5$ $\Omega^{-1}$m$^{-1}$). Figure S1(b) shows the representative temperature dependence of $\sigma$ for a representative 10 nm-thick Bi$_{0.92}$Sb$_{0.08}$ thin film. $\sigma$ shows metallic behavior as observed in various thin BiSb films, which reflects the conduction in the surface states.[17]

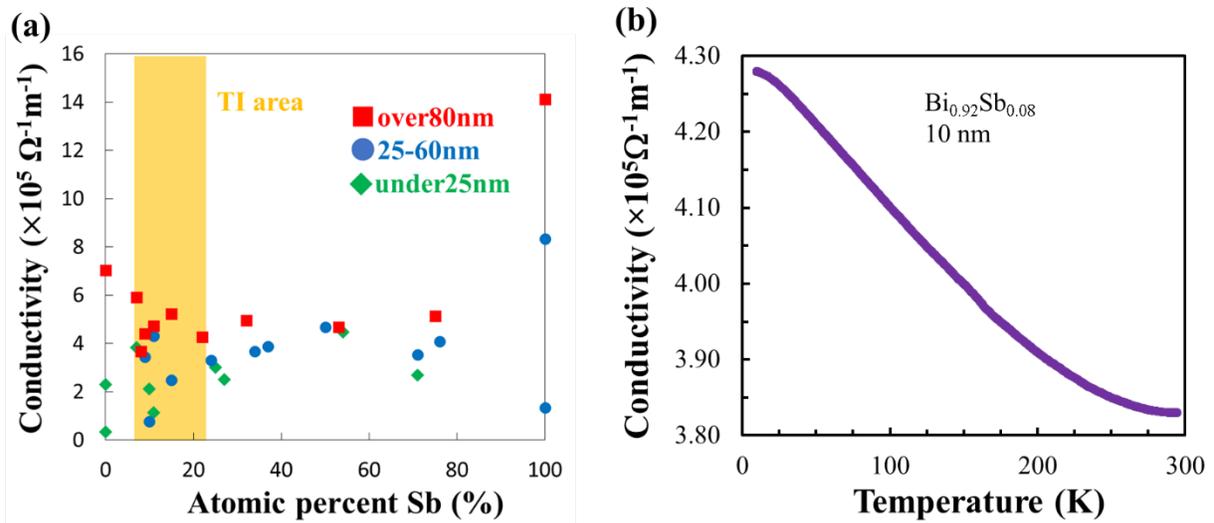

**Figure S1. Conductivity of BiSb thin films.** (a) Conductivity $\sigma$ of BiSb thin films at various thickness and Sb atomic concentration at 270 K (Ref.17). (b) Temperature dependence of $\sigma$ for a representative 10 nm-thick Bi$_{0.92}$Sb$_{0.08}$ thin film.

## 2. RHEED patterns during growth of the BiSb / MnGa bi-layers

Figure S2 shows the RHEED patterns during growth of the BiSb / MnGa bi-layers. The RHEED patterns of both MnGa (3) and BiSb (10) are streaky, indicating good crystal quality. Particularly, the interface between MnGa and BiSb is atomically flat, as seen in Fig. S2(c) and S2(d).

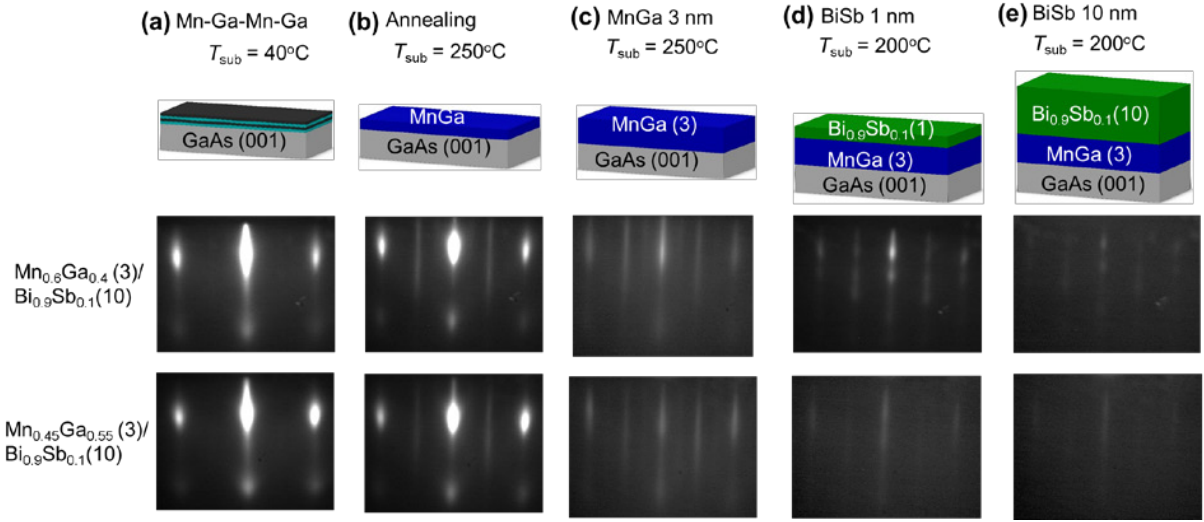

**Figure S2. RHEED patterns during growth of the BiSb / MnGa bi-layers.** (a) After deposition of the 4 monolayer Mn-Ga-Mn-Ga template, (b) after annealing the template at 250°C, (c) after growth of the 3 nm-thick MnGa layer, (d) after growth of a 1 nm-thick BiSb layer at 200°C, and (e) after growth of the 10 nm-thick BiSb layer.

To check the crystallographic orientation of the BiSb / MnGa bi-layers, we grew a thick BiSb (20) / MnGa (10) bi-layer, and performed X-ray diffraction measurements. Figure S3 shows the $\theta$-$2\theta$ XRD spectrum of the BiSb (20) / MnGa (10) bi-layer. The XRD spectrum indicates that the crystallographic orientation is BiSb(012)//MnGa(001)//GaAs(001).

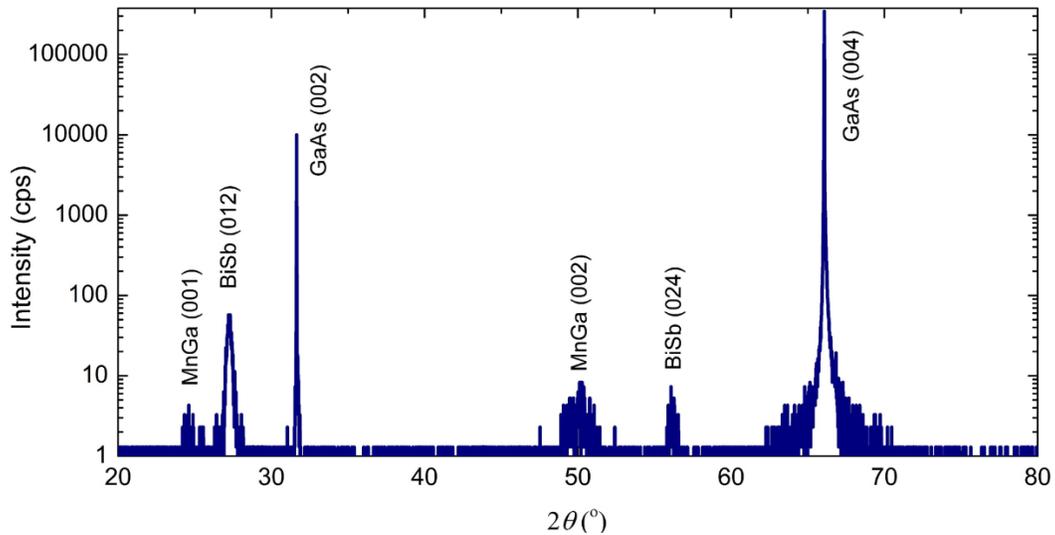

**Figure S3. $\theta$-$2\theta$ XRD spectrum of a BiSb (20) / MnGa (10) bi-layer.** The crystallographic orientation is BiSb(012)//MnGa(001)//GaAs(001).

## 3. Absence of the field-like spin-orbit field

In addition to the damping-like spin-orbit field $H_{so}$ described in the manuscript, there is possibly a field-like spin-orbit field $H_T$ generated by BiSb. In this section, we will show that $H_T$ is negligible by analyzing the magnitude of the $R_H$ loops of the $Bi_{0.9}Sb_{0.1}$ (10) / $Mn_{0.45}Ga_{0.55}$ (3) Hall bar under a slightly tilting perpendicular magnetic field ($\theta = 2°$ toward the $x$ direction) (full data set are shown in Fig. 2(b) of the manuscript). Here, we consider a situation when $H_T$ was comparable with $H_{so}$, i.e. its strength was ~ 1 kOe at the highest $J = 13.8\times10^5$ A/cm$^2$. Since $H_T$ is along the $y$ direction, it can generate a non-zero $M_y$ component for the $Mn_{0.45}Ga_{0.55}$(3) layer. Therefore, there would be contribution from the planar Hall resistance $R_{PHE}$ in addition to the anomalous Hall resistance $R_{AHE}$ to the total Hall resistance $R_H$. Unlike $H_{so}$, however, the direction of $H_T$ depends only on the direction of $J$, but not $M$. Figures S4(a) and S4(b) show the direction of $M_x$, $M_y$, and $M_z$ at a perpendicular $H_{ext} = 8$ kG and -8 kG, respectively. In Fig. S4(a), $M_x > 0$, $M_y > 0$, and $M_z > 0$, thus $R_H(8$ kG$) = R_{PHE}*\sin(2\phi) + R_{AHE}$, where $\phi = \tan^{-1}(M_y/M_x)$. In Fig. S4(b), $M_x < 0$, $M_y > 0$, and $M_z < 0$, thus $R_H(-8$ kG$) = R_{PHE}*\sin[2(\pi-\phi)] - R_{AHE} = -R_{PHE}*\sin(2\phi) - R_{AHE}$. That means the magnitude of the $R_H$ loop is given by $R_{PHE}*\sin(2\phi) + R_{AHE}$, and would be enhanced by the amount of $R_{PHE}*\sin(2\phi)$. As a result, one would expect that the magnitude of the $R_H$ loops would systematically increase with increasing $J$. However, as seen in Fig. 2(b) of the manuscript, the magnitude of the $R_H$ loops does not change when $J$ is increased from $0.2\times10^5$ A/cm$^2$ to $13.8\times10^5$ A/cm$^2$. This indicates that $H_T$ is negligible, and that $R_H$ has only contribution from $R_{AHE}$.

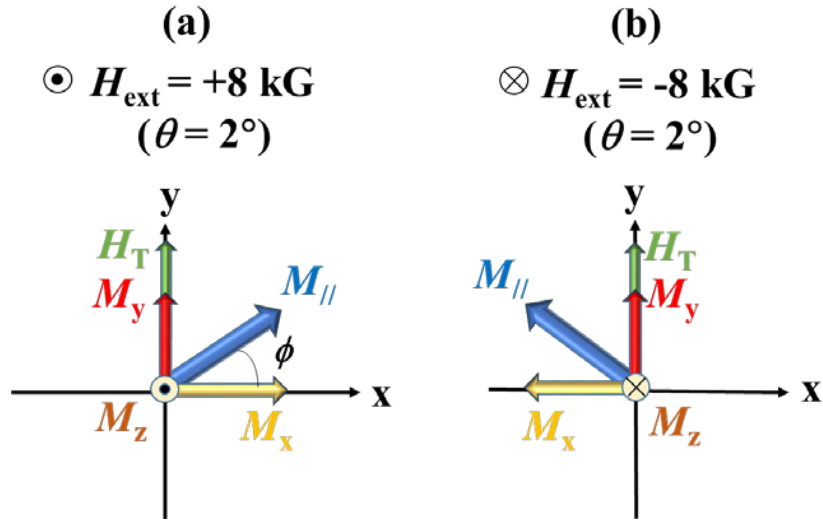

**Figure S4.** Direction of $M_x$, $M_y$, and $M_z$ under a slightly tilting perpendicular magnetic field $H_{ext}$ ($\theta = 2°$ toward the $x$ direction), assuming a non-zero field-like spin-orbit field $H_T$. **(a)** When $H_{ext}$ = 8 kG, and **(b)** when $H_{ext}$ = -8 kG, respectively.